\documentclass[showpacs, secnumarabic, amssymb, nobibnotes, aps, prb, reprint]{revtex4-1}
\usepackage{graphicx}%
\usepackage{dcolumn}%
\usepackage{bm}%
\usepackage{lineno}

\begin{document}

\title{The magnetic interactions in spin-glasslike
Ge$_{1\textrm{-}x\textrm{-}y}$Sn$_{x}$Mn$_{y}$Te \\ diluted magnetic
semiconductor}

\author{L.~Kilanski}
\email[Electronic mail:]{kilan@ifpan.edu.pl}
\author{R.~Szymczak}
\author{W.~Dobrowolski}
\affiliation{Institute of Physics, Polish Academy of Sciences, al.
Lotnikow 32/46, 02-668 Warsaw, Poland}

\author{K.~Sza{\l}owski}
\affiliation{Department of Solid State Physics, University of
{\L}\'{o}d\'{z}, ul. Pomorska 149/153, 90-236 {\L}\'{o}d\'{z},
Poland}

\author{V.~E.~Slynko}
\author{E.~I.~Slynko}
\affiliation{Institute of Materials Science Problems, Ukrainian
Academy of Sciences, 5 Wilde Street, 274001 Chernovtsy, Ukraine}

\date{\today}

\begin{abstract}

We investigated the nature of the magnetic phase transition in
Ge$_{1\textrm{-}x\textrm{-}y}$Sn$_{x}$Mn$_{y}$Te mixed crystals with
chemical composition changing in the range of
0.083$\,$$\leq$$\,$$x$$\,$$\leq$$\,$0.142 and
0.012$\,$$\leq$$\,$$y$$\,$$\leq$$\,$0.119. The DC magnetization
measurements performed in magnetic fields up to 90$\;$kOe and
temperature range 2$\,$$\div$$\,$200$\;$K showed that the magnetic
ordering at temperatures below $T$$\,$$=$$\,$50$\;$K exhibits
features characteristic for both spin-glass and ferromagnetic
phases. The modified Sherrington - Southern model was applied to
explain the observed transition temperatures. The calculations
showed that the spin-glass state is preferred in the range of the
experimental carrier concentrations and Mn contents. The value of
the Mn hole exchange integral was estimated to be
$J_{pd}$$\,$$=$$\,$0.45$\pm$0.05$\;$eV. The experimental
magnetization vs temperature curves were reproduced satisfactorily
using the non-interacting spin-wave theory with the exchange
constant $J_{pd}$ values consistent with those calculated using
modified Sherrington - Southern model. The magnetization vs magnetic
field curves showed nonsaturating behavior at magnetic fields
$B$$\,$$<$$\,$90$\;$kOe indicating the presence of strong magnetic
frustration in the system. The experimental results were reproduced
theoretically with good accuracy using the molecular field
approximation-based model of a disordered ferromagnet with
long-range RKKY interaction.

\end{abstract}





\maketitle


\section{Introduction}

The IV-VI group of ferromagnetic compounds e.g.
Ge$_{1\textrm{-}x}$Mn$_{x}$Te possesses unique possibility to tune
the magnetic and electrical properties of the crystals independently
[\onlinecite{Fukuma02a}]. The magnetism of bulk (Ge,Mn)Te crystals
was for the first time studied over 40 years ago by Rodot et. al
[\onlinecite{Rodot66a}]. The ferromagnetic ordering observed in the
Ge$_{1\textrm{-}x}$Mn$_{x}$Te with the Curie temperatures $T_{C}$ as
high as 167$\;$K [\onlinecite{Cochrane74a}] for the crystals with
the composition $x$$\,$$=$$\,$0.5 was attributed to the indirect
long range Ruderman-Kittel-Kasuya-Yosida (RKKY) interaction.
Recently, molecular beam epitaxy (MBE) grown
Ge$_{1\textrm{-}x}$Mn$_{x}$Te thin layers are the subject of great
scientific interest\cite{Fukuma06a, Chen06a, Lechner07a, Knoff08a,
Lim09a, Kowalski10a}. The progress in the methods of growth of this
compound allowed the increase of the observed Curie temperatures to
values as high as 200$\;$K [\onlinecite{Lechner07a}] and gives hope
to obtain the room temperature ferromagnetism. \\ \indent In this
paper we extend our previous studies of dynamic magnetic properties
of \linebreak Ge$_{1\textrm{-}x\textrm{-}y}$Sn$_{x}$Mn$_{y}$Te
crystals [\onlinecite{Kilanski08a}, \onlinecite{Kilanski09a}]. The
results of the measurements of the AC susceptibilities showed the
occurrence of the magnetic transitions into the spin-glasslike state
at temperatures lower than 50$\;$K. In this paper we present the
measurements of the static magnetic properties i.e. the
magnetization measurements in the presence of a static magnetic
field. \\ \indent The observed magnetic properties are discussed in
theoretical context. The Sherrington-Southern model, capturing the
essentials of spin-glass physics, allows us to predict the relevant
transition temperatures and to estimate the value of the exchange
integral $J_{pd}$ in the studied crystals, assuming the RKKY
interaction between magnetic impurities. The temperature dependence
of magnetization will be modeled using the noninteracting spin wave
theory, including the same RKKY interaction, which supports the
estimates of $J_{pd}$. To describe the behavior of the high-field
magnetization curves, we construct a molecular field
approximation-based model of a disordered ferromagnet with
long-range RKKY interaction, putting special emphasis on the
possible magnetic inhomogenity in the system, caused by presence of
antiferromagnetic superexchange interaction between nearest-neighbor
magnetic ions. This model is used to explain the non-saturating
behavior of magnetization.

\section{Sample preparation and basic characterization}

For purposes of this research bulk
Ge$_{1\textrm{-}x\textrm{-}y}$Sn$_{x}$Mn$_{y}$Te crystals were
prepared using modified Bridgman method. The growth method was
modified according to the ideas proposed by Aust and Chalmers for
the alumina crystals \cite{Aust58a}. The radial temperature gradient
was present in the growth furnace, creating 15$\;$deg slope of the
crystallization front. It allowed the decrease of the crystal blocks
in the ingot from a few down to a single one. \\ \indent The
chemical composition of the alloy was determined using the x-ray
fluorescence method (with typical relative uncertainty of about
10\%). The as grown ingots were cut into 1$\;$mm thick slices
perpendicular to the growth direction. The measured chemical content
changed continuously along the growth direction. For the present
investigations we used the very same samples as presented in
Ref.$\;$[\onlinecite{Kilanski09a}] i.e. the studied crystals had
chemical composition changing in a range of values
0.083$\,$$\leq$$\,$$x$$\,$$\leq$$\,$0.142 and
0.012$\,$$\leq$$\,$$y$$\,$$\leq$$\,$0.119. The crystal slices
selected for the present investigations had small inhomogeneity of
the chemical composition of the order of the uncertainty of x-ray
fluorescence method i.e. less than 10\% of the resultant molar
fraction $x$ \\ \indent The crystallographic quality of
Ge$_{1\textrm{-}x\textrm{-}y}$Sn$_{x}$Mn$_{y}$Te crystals was
studied by means of the x-ray diffraction (XRD). Analysis of the XRD
measurement results showed that the studied crystals were single
phased. They crystallized in the [111] direction distorted NaCl i.e.
the rhombohedral structure. The lattice parameters obtained for the
studied crystals had values similar to those reported in the
literature for germanium telluride bulk crystals e.g. the lattice
parameter $a$ was around 5.98$\;$$\textrm{\AA}$ and the angle of
distortion $\alpha$ was around 88.3$\;$[deg]
[\onlinecite{Galazka99a}]. \\ \indent Basic electrical properties of
Ge$_{1\textrm{-}x\textrm{-}y}$Sn$_{x}$Mn$_{y}$Te crystals were also
characterized. The measurements (using the standard DC six contact
technique) of the resistivity parallel to the current direction
$\rho_{xx}$ and the Hall effect (using constant magnetic field
$B$$\,$$=$$\,$14$\;$kOe) for each sample at room temperature were
performed. The samples had parallelepiped shape with typical
diameters around 1$\times$1$\times$8 mm. The results showed that the
studied crystals were highly degenerated $p$-type semiconductors
with high hole concentration $n$$\,$$>$$\,$10$^{21}$$\;$cm$^{-3}$
and low carrier mobilities $\mu$$\,$$\leq$$\,$100$\;$cm$^{2}$/(Vs).
For more detailed discussion of electrical properties of
Ge$_{1\textrm{-}x\textrm{-}y}$Sn$_{x}$Mn$_{y}$Te see
Refs.~\onlinecite{Kilanski08a} and \onlinecite{Kilanski09a}.

\section{Results and discussion}

The DC magnetometry was used in the present studies of the
Ge$_{1\textrm{-}x\textrm{-}y}$Sn$_{x}$Mn$_{y}$Te crystals. The very
same sample pieces as in the case of transport characterization with
electrical contacts removed were studied by means of DC
magnetometry. The magnetization measurements were performed using
two magnetometers. The temperature dependent magnetization in a
small magnetic fields was collected using a Quantum Design SQUID
MPMS XL-5 magnetometer. The isothermal magnetization curves and
hysteresis loops were measured using the extraction method employed
by the LakeShore 7229 DC magnetometer. This method has the lower
accuracy than SQUID magnetometer, but allowed us to measure at
magnetic fields as high as $B$$\,$$=$$\,$90$\;$kOe.

\subsection{Low field magnetization}

\begin{figure}[!h]
  \begin{center}
    \includegraphics[width = 0.5\textwidth, bb = 16 30 278 250]
    {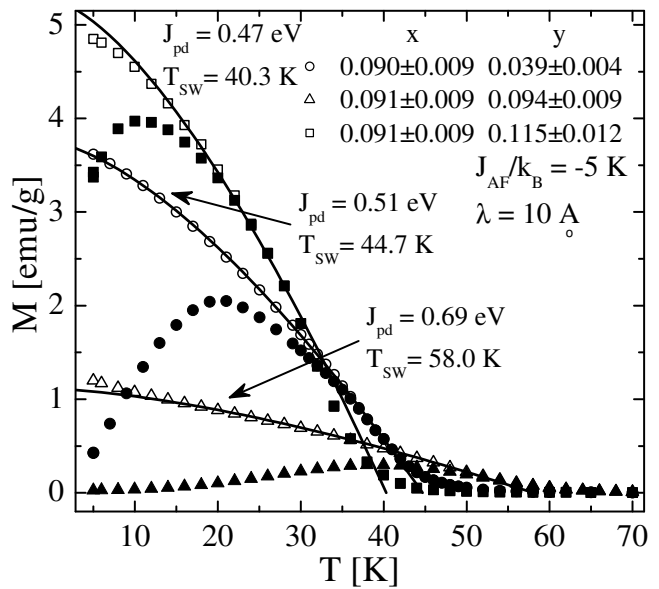}\\
  \end{center}
  \caption{\small Experimental (points) and theoretical (lines)
  magnetization as a function of temperature for
  samples cooled in the absence (closed symbols) and the presence
  (open symbols) of external magnetic
  field $B$$\,$$=$$\,$50$\;$Oe for two selected
  Ge$_{1\textrm{-}x\textrm{-}y}$Sn$_{x}$Mn$_{y}$Te samples with
  different chemical composition (labels).}
  \label{fig01}
\end{figure}

The temperature dependencies of the zero-field-cooled (ZFC) and
field-cooled (FC) magnetization were measured in the temperature
range between 4.3 and 100 K using the constant magnetic field
$B$$\,$$=$$\,$50$\;$Oe. Typical results of the temperature
dependencies of the magnetization for selected
Ge$_{1\textrm{-}x\textrm{-}y}$Sn$_{x}$Mn$_{y}$Te crystals with
constant Sn and Mn contents varied between 0.039 and 0.115 are
presented in Figure$\;$\ref{fig01}.
The differences between the FC and ZFC curves for the studied
crystals were observed. Bifurcations between ZFC and FC
magnetization values at low temperatures are typical features of the
appearance of a spin-glass phase. \\ \indent The
paramagnet--spin-glass transition temperatures $T_{SG}$ were
determined from the temperature dependent magnetization curves at
the bifurcation points. The values obtained in this way are in good
agreement with our previously determined $T_{SG}$ values from AC
magnetometry results (see Ref.$\;$\onlinecite{Kilanski09a}). Both
the dynamic and static magnetic properties of
Ge$_{1\textrm{-}x\textrm{-}y}$Sn$_{x}$Mn$_{y}$Te crystals show
features similar to those observed in canonical spin-glass systems
e.g. CuMn or AuFe diluted metallic alloys\cite{Mydosh94a}. \\
\indent The presented experimental FC $M(T)$ curves for weak field
were theoretically reproduced with the use of the non-interacting
spin wave theory. The magnon dispersion relation has been assumed in
the isotropic form $\epsilon\left(k\right)=Dk^2$, with spin-wave
stiffness coefficient
$D=\frac{1}{6}yS\sum_{i}^{}{J\left(R_{ij}\right)R^{2}_{ij}}$, where
$J\left(R_{ij}\right)$ is the RKKY exchange integral, given by a
formula (\ref{Eq03}), calculated for a pair of ions in the lattice
with a mutual distance of $R_{ij}$. In the present considerations it
is assumed that the indirect RKKY interaction with the exchange
constant $J_{RKKY}$ has the following form\cite{Ruderman54a,
Kasuya56a, Yoshida57a}
\begin{widetext}
\begin{equation}\label{Eq03}
J_{RKKY}(R_{ij})=N_{V}\frac{m^{*}J_{pd}^{2}a_{s}^{6}k_{F}^{4}}{32\pi^{3}\hbar^{2}}
\times \frac{\sin(2k_{F}R_{ij})-2k_{F}R_{ij}\cos(2k_{F}R_{ij})}{(2k_{F}R_{ij})^{4}}\,\exp(-\frac{R_{ij}}{\lambda}),
\end{equation}
\end{widetext}
where $m^{*}$ is the effective mass of the carriers, $J_{pd}$ is the
exchange integral between conducting holes and Mn ions, $a_{0}$ is
the lattice parameter, $k_{F} = (3 \pi^{2} n / \hbar N_{V})$ is the
Fermi wave vector, $N_{V}$ is the number of the valleys of the
valence band, $\hbar$ is the Planck constant divided per 2$\pi$, and
$R_{ij}$ is the distance between magnetic ions. For nearest-neighbor
ions, the additional antiferromagnetic superexchange coupling,
parameterized by $J_{AF}$, has been included. This results from
averaging the magnon energies over possible orientations of
wavevector $\vec{k}$ for its given length $k$ (similar to the
calculation in Ref.~\onlinecite{Cochrane74a}). \\ \indent We
emphasize  that using such a direction-averaged dispersion relation
for our system is motivated by the presence of spatial disorder in a
diluted magnetic system, which makes the influence of the exact
lattice structure (involved in calculating the Fourier transform)
less pronounced. Moreover, we limit our considerations to a
quadratic part of the dispersion relation instead of using the full
expression valid for the whole Brillouin zone. This reflects the
fact that in a strongly diluted system the typical distances between
the magnetic ions are noticeably larger than the lattice constant
itself, so that only low-energy spin waves (of wavelength exceeding
the mentioned distance) constitute well-defined excitations. The
mentioned assumptions lead tho the temperature dependence of
magnetization following exactly the Bloch's law in the form
\begin{equation}\label{Eq04}
    M\left(T\right)=M\left(0\right)\left[1-\left(T/T_{SW}\right)^{3/2}\right].
\end{equation}
The characteristic temperature $T_{SW}$ can be expressed as
\begin{equation}\label{Eq05}
T_{SW}=\frac{2\pi Da^2}{k_{B}} \left(\frac{4S}{\zeta\left(3/2\right)}\right)^{2/3},
\end{equation}
where $a$ is
the lattice constant and $\zeta\left(3/2\right)\simeq 2.612$ is the
appropriate Riemann zeta function. The performed calculations
allowed us to satisfactorily reproduce the experimentally observed
$M(T)$ curves (see Fig.$\;$\ref{fig01}) for the $J_{pd}$ exchange
coupling constant values consistent with those obtained from the
further analysis of Curie-Weiss and spin-glass transition
temperatures within the Sherrington-Southern model. It is worth
mentioning that such a robust $T^{3/2}$-like behavior of $M(T)$
curves has also been noticed by [\onlinecite{Cochrane73a}] and
[\onlinecite{Fukuma08b}]. \\ \indent The proposed theoretical
approach was not able to take fully into account the magnetic
frustration in the studied system, which is reflected in the shape
(long tail) of the $M(T)$ curve at $T$$\,$$\simeq$$\,$$T_{SG}$.

\subsection{Estimation of the exchange integral $J_{pd}$}

\begin{figure}[t]
  \begin{center}
    \includegraphics[width = 0.5\textwidth, bb = 16 30 270 250]
    {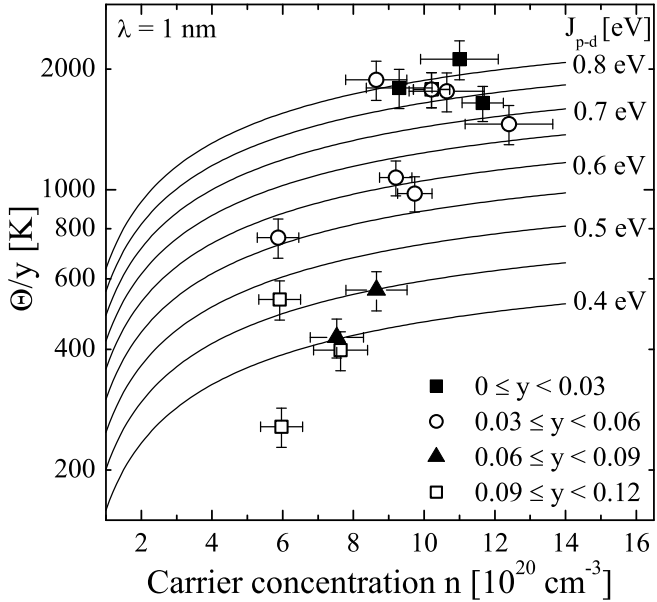}\\
  \end{center}
  \caption{\small The Curie-Weiss temperature normalized to the Mn
  molar fraction y as a function of the carrier concentration $n$
  calculated theoretically (lines) using values of the
  exchange integral $J_{pd}$ varying between 0.4$\;$eV to 0.8$\;$eV
  with step 0.05$\;$eV, and the experimental values (points) obtained
  for the studied Ge$_{1\textrm{-}x\textrm{-}y}$Sn$_{x}$Mn$_{y}$Te
  crystals grouped with respect to the amount of Mn ions in the alloy
  $y$ (see legend).} \label{fig02}
\end{figure}

\begin{figure}[t]
  \begin{center}
    \includegraphics[width = 0.5\textwidth, bb = 0 55 860 850]
    {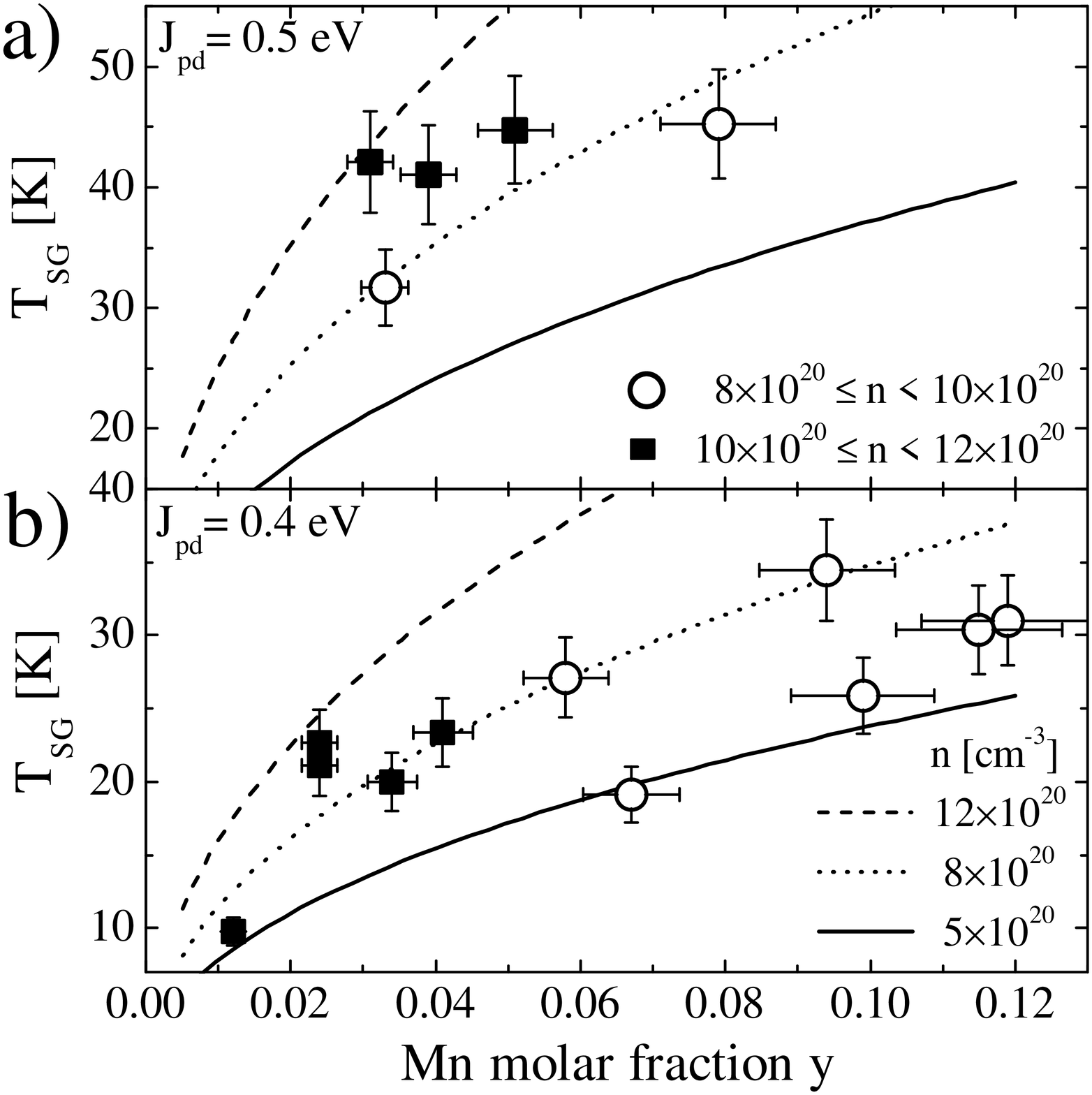}\\
  \end{center}
  \caption{\small The spin glass transition temperatures $T_{SG}$ as a
  function of the Mn molar fraction $y$ calculated (lines) for two
  different values of the exchange integral $J_{pd}$ and different
  values of carrier concentration $n$ (see legend), and the
  experimentally determined values of $T_{SG}$ for
  Ge$_{1\textrm{-}x\textrm{-}y}$Sn$_{x}$Mn$_{y}$Te crystals with
  different chemical content $y$. The points were grouped with respect
  to the different carrier concentrations (see legend).}
  \label{fig03}
\end{figure}

The spin-glass state in the diluted magnetic material is well
described by the model proposed by Sherrington and
Southern\cite{Sherrington75b}. The Sherrington--Southern (SS) model
was found to be the most appropriate approach for the description of
systems with a large number of neighboring paramagnetic ions.
Moreover, the modified SS model was applied to compounds belonging
to the group of IV-VI semimagnetic semiconductors i.e.
Sn$_{1\textrm{-}x}$Mn$_{x}$Te and
Pb$_{1\textrm{-}x\textrm{-}y}$Sn$_{x}$Mn$_{y}$Te
[\onlinecite{Eggenkamp95a}]. In the present work we adapted the
modified SS model [\onlinecite{Eggenkamp95a}] to quantify the
$J_{pd}$ exchange interaction in the studied
Ge$_{1\textrm{-}x\textrm{-}y}$Sn$_{x}$Mn$_{y}$Te mixed crystals.
Sherrington and Southern proposed the Gaussian distribution of the
interaction strength in Heisenberg Hamiltonian, with mean value
$J_{0}$ and width $\Delta J$. The model allows the calculations of
both paramagnetic Curie-Weiss $\Theta$ and paramagnet--spin-glass
transition $T_{SG}$ temperatures given by Eqs.$\;$\ref{Eq02a} and
\ref{Eq02b} {\setlength\arraycolsep{5pt}
\begin{eqnarray}
  \Theta &=& \frac{2S(S+1)x}{3k_{B}} J_{0},  \label{Eq02a}\\
  T_{SG} &=& \frac{2 \Delta J}{3 k_{B}} \big{[} S^{2}(S+1)^{2}
  + S(S+1)/2  \big{]}^{1/2},  \label{Eq02b}
\end{eqnarray}}
\\ \noindent where $k_{B}$ is the Boltzmann constant. In the above
equations the RKKY indirect interactions has the exchange constant
$J_{RKKY}$ is expressed using the equation~\ref{Eq03}. We assumed
that the band structure of the studied
Ge$_{1\textrm{-}x\textrm{-}y}$Sn$_{x}$Mn$_{y}$Te crystals is similar
to that of GeTe. GeTe has got a nearly spherical valence band with
maximum at the $L$-point of the Brillouin zone\cite{Lewis69a} the
effective mass of the carriers in the four-fold degenerated $L$-band
equal to 1.2$\cdot$$m_{e}$ [\onlinecite{Lewis73a}], where $m_{e}$ is
the free electron mass, as in the case of the GeTe crystals. The SS
model assumes the finite range of the interactions by introducing
the exponential damping factor $\lambda$. In our calculations we
assumed $\lambda$ to be equal to 1$\;$nm what is a reasonable value
for the semiconductor with relatively low carrier mobility
$\mu$$\,$$<$$\,$100$\;$cm$^{2}$/(Vs) (see Refs.
\onlinecite{Kilanski08a} and \onlinecite{Kilanski09a}). The lattice
parameters for the pure GeTe crystals were used in our calculations.
\\ \indent The results of calculation of the $\Theta(n)$
dependencies performed for values of the Mn-hole exchange integral
$J_{pd}$ varying between 0.4 and 0.8 eV are presented in
Fig.$\;$\ref{fig02}.
Most of the experimental points lie between the $\Theta(n)$ curves
calculated for $J_{pd}$ varying between the two values
0.4$\,$$\div$$\,$0.8$\;$eV. Because of the large scatter of the
experimental values we found it reasonable to focus on the
spin-glass transition temperature $T_{SG}$ estimations which as will
be seen give more accurate results. \\ \indent The $T_{SG}$
dependence on the Mn content for a few values of the hole
concentrations is presented in Fig.$\;$\ref{fig03}.
The experimentally determined values of $T_{SG}(y)$ are well
described by the theoretical curves with the values of the exchange
constant $J_{pd}$ changing between 0.4 and 0.5$\;$eV. It must be
noted, that for the samples for which the $T_{SG}$ values are in the
vicinity of $J_{pd}$$\,$$=$$\,$0.5$\;$eV, the coercive fields were
the highest in the entire series of
Ge$_{1\textrm{-}x\textrm{-}y}$Sn$_{x}$Mn$_{y}$Te crystals. It may be
interpreted that the estimated higher value of the exchange constant
for these crystals is not an artifact but reflects their magnetic
properties. \\ \indent The estimated values of $J_{pd}$ for the
studied Ge$_{1\textrm{-}x\textrm{-}y}$Sn$_{x}$Mn$_{y}$Te crystals
are within the range between reported in literature for
Sn$_{1\textrm{-}x}$Mn$_{x}$Te bulk crystals
[\onlinecite{Eggenkamp94a}] ($J_{pd}$$\,$$\approx$$\,$0.1$\;$eV for
$x$$\,$$\leq$$\,$0.1) and Ge$_{1\textrm{-}x}$Mn$_{x}$Te
($J_{pd}$$\,$$\approx$$\,$0.8$\,$$\div$$\,$0.9$\;$eV for
$x$$\,$$>$$\,$0.1) bulk crystals [\onlinecite{Cochrane73a},
\onlinecite{Cochrane74a}], and $J_{pd}$$\,$$\approx$$\,$0.7$\;$eV
for $x$$\,$$=$$\,$0.07 as obtained in Ge$_{1\textrm{-}x}$Mn$_{x}$Te
epitaxial layers [\onlinecite{Fukuma03a}].

\subsection{High field magnetization}

\begin{figure*}[t]
  \begin{center}
    \includegraphics[width = 0.9\textwidth, bb = 10 30 345 194]
    {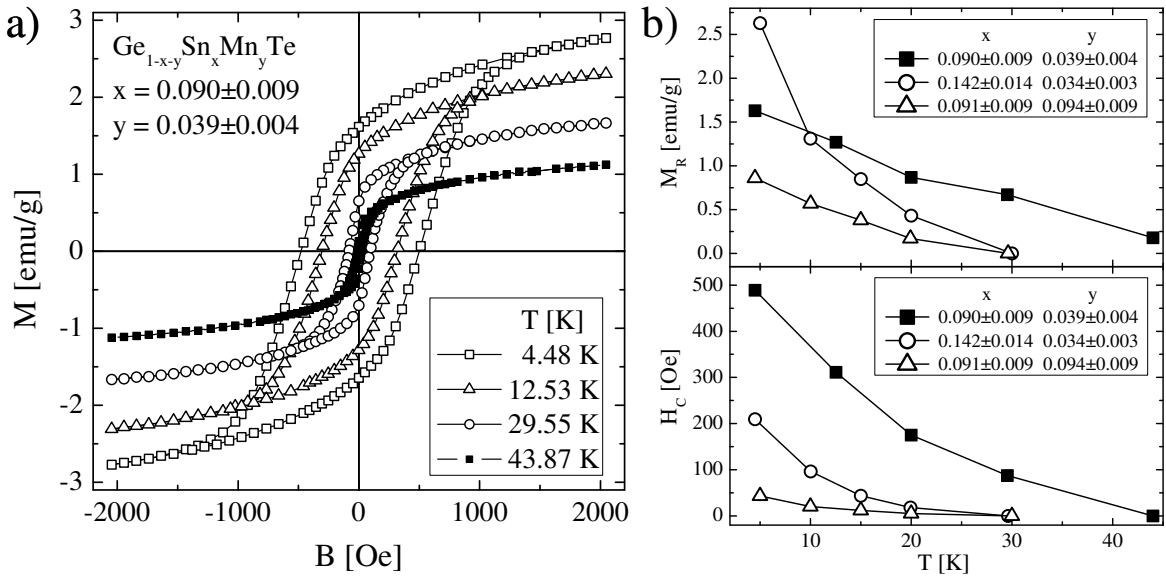}\\
  \end{center}
  \caption{\small a) Hysteresis loops measured at selected constant
  temperatures (see legend) for selected
  Ge$_{1\textrm{-}x\textrm{-}y}$Sn$_{x}$Mn$_{y}$Te sample (chemical
  composition shown in legend) and b) spontaneous magnetization
  $M_{R}$ and coercive field $H_{C}$ as a function of temperature
  for a few selected Ge$_{1\textrm{-}x\textrm{-}y}$Sn$_{x}$Mn$_{y}$Te
  samples with different chemical composition (see legend).}
  \label{fig04}
\end{figure*}

\begin{figure}[t]
  \begin{center}
    \includegraphics[width = 0.5\textwidth, bb = 15 31 280 250]
    {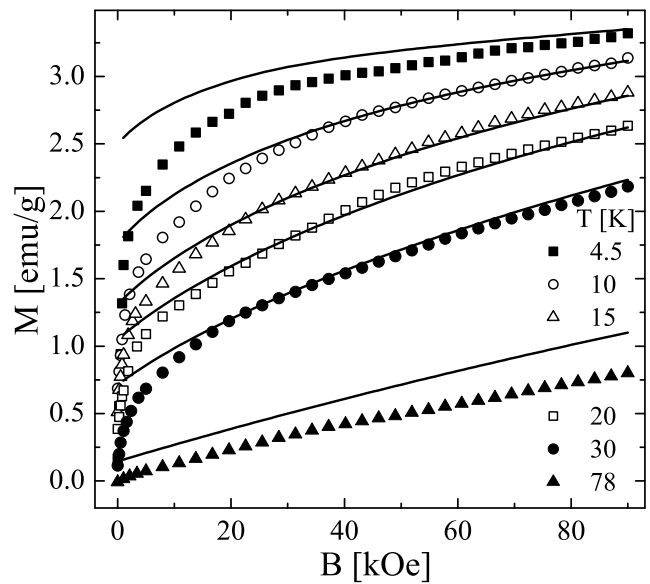}\\
  \end{center}
  \caption{\small Isothermal magnetization curves $M(B)$ measured
  (points) and reproduced using the mean field model described in the
  Appendix (lines) for the parameters
  $J_{pd}$$\,$$=$$\,$0.75$\;$eV, $y$$\,$$=$$\,$0.034$\;$,
  $p$$\,$$=$$\,$0.71$\;$, $\lambda$$\,$$=$$\,$15$\;$\AA,
  $J_{AF}/k_{\rm B}$$\,$$=$$\,$-20$\;$K at different temperatures (see
  legend) for selected
  Ge$_{1\textrm{-}x\textrm{-}y}$Sn$_{x}$Mn$_{y}$Te sample with
  $x$$\,$$=$$\,$0.142$\pm$0.014 and $y$$\,$$=$$\,$0.034$\pm$0.003.}
  \label{fig05}
\end{figure}

\begin{figure}[t]
  \begin{center}
    \includegraphics[width = 0.5\textwidth, bb = 10 30 280 267]
    {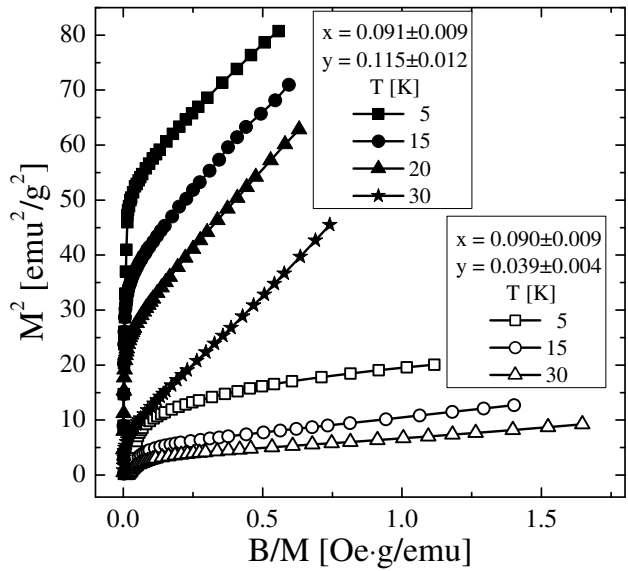}\\
  \end{center}
  \caption{\small The Arrot plot for two selected
  Ge$_{1\textrm{-}x\textrm{-}y}$Sn$_{x}$Mn$_{y}$Te
  crystals with different chemical composition (see legend).}
  \label{fig06}
\end{figure}

The magnetic properties of
Ge$_{1\textrm{-}x\textrm{-}y}$Sn$_{x}$Mn$_{y}$Te crystals is the
presence of high magnetic field for $B$$\,$$\leq$$\,$90$\;$kOe were
studied at temperatures between
4.3$\,$$\leq$$\,$$T$$\,$$\leq$$\,$80$\;$K. At magnetic fields
$B$$\,$$\leq$$\,$600$\;$Oe an irreversible behavior with clear
hysteresis is observed. Typical results of hysteresis curves
recorded for a Ge$_{0.871}$Sn$_{0.090}$Mn$_{0.039}$Te sample at a
few temperatures are presented in Fig.$\;$\ref{fig04}.
The magnetic hysteresis is not a typical feature in canonical
spin-glasses such as CuMn [\onlinecite{Mydosh94a}] but was observed
for many spin-glass systems such as AuFe with 8\% at. Fe
[\onlinecite{Prejean80a}] or NiMn with 21\% at. Mn
[\onlinecite{Senoussi84a}].  \\ \indent The analysis of hysteresis
loops showed that there exists a correlation between the Mn content
$y$, the values of spontaneous magnetization $M_{R}$, and the
coercive field $H_{C}$ (see Fig.$\;$\ref{fig04}b). When $y$ changes
between 0.012 and 0.039 the values (measured at
$T$$\,$$\approx$$\,$4.5$\;$K) of $M_{R}$ increase from 0.45$\;$emu/g
to 2.5$\;$emu/g and $H_{C}$ values are increasing from 80$\;$Oe to
430$\;$Oe. In the case of the group of crystals with
$y$$\,$$>$$\,$0.039 the inverse dependencies were observed, namely,
the reduction of both $M_{R}$ and $H_{C}$ with increasing $y$. In
the case of the group of crystals with Mn content $\approx$0.035 and
Sn content $x$ changing between $x$$\,$$=$$\,$0.090 and 0.145, the
changes of the $H_{C}$ and $M_{R}$ can be attributed to the increase
of the carrier concentration. The coercive field $H_{C}$ becomes
smaller with increasing amount of tin ions changing from 430$\;$Oe
to 210$\;$Oe for crystals with $x$$\,$$=$$\,$0.090 and 0.145. In
turn, the value of $M_{R}$ showed the opposite trend rising from 1.6
to 2.6$\;$emu/g. \\ \indent The observed parameters of the magnetic
hysteresis loop and their changes with the chemical composition of
the alloy can be associated with the modification of the domain
structure of the material. Furthermore, together with the increasing
Mn content the probability for antiferromagnetic pairing of Mn ions
increases. \\ \indent The $H_{C}$ values observed in our studies are
smaller than reported in the literature for bulk
Ge$_{1\textrm{-}x}$Mn$_{x}$Te crystals [\onlinecite{Rodot66a}]
($H_{C}$$\,$$<$$\,$1200$\;$Oe for $x$$\,$$=$$\,$0.048), but are
close to those reported for thin Ge$_{1\textrm{-}x}$Mn$_{x}$Te
epitaxial layers [\onlinecite{Fukuma03a}]
($H_{C}$$\,$$<$$\,$700$\;$Oe for $x$$\,$$<$$\,$0.6). \\ \indent The
typical results for the Ge$_{0.824}$Sn$_{0.142}$Mn$_{0.034}$Te
crystal obtained in the range of magnetic fields up to 90$\;$kOe and
temperatures lower than 80$\;$K are presented in
Fig.$\;$\ref{fig05}.
The $M(B)$ curves do not saturate and are nearly linear in the range
of magnetic fields $B$$\,$$\geq$$\,$30$\;$kOe. The lack of
saturation of $M$ reflects the fact that the studied alloy is not a
ferromagnet but a spin glass at the temperatures below 50$\;$K \\
\indent In order to model the field dependence of magnetization, we
performed calculations within the molecular-field approximation,
constructing the model described in the Appendix. The numerical
calculations were performed for the exchange constant $J_{pd}$, the
Mn amount $y$, the fraction of isolated magnetic ions $p$, the
interaction decay distance $\lambda$, and the energy of the
antiferromagnetic interactions in the system $J_{AF}$ specified and for 2$\cdot$10$^{4}$
realizations of disorder, summing the interactions between the
magnetic ions up to the distance of 100 \AA. The method allowed to
satisfactorily reproduce the high-field behavior of $M(H)$ curves,
i.e. the linear increase of magnetization with the field. It must be
emphasized that the method adopted is based on the molecular-field
approximation, so that it cannot be expected to account for the
collective excitations in the system and to yield the precise
temperature dependence of magnetization. It is obvious that the
low-field magnetic susceptibility is underestimated by the present
model, which might be attributed to neglect of correlations. \\
\indent In Fig.$\;$\ref{fig06} the isothermal magnetization curves
obtained for the studied crystals are presented in the form of Arrot
plots.
With the increase of the Mn content in the samples the experimental
$M^{2}(B/M)$ curves change their character. Such a behavior is
usually attributed to the change of the strength of the
ferromagnetic interactions in the system. In the case of the
Ge$_{1\textrm{-}x\textrm{-}y}$Sn$_{x}$Mn$_{y}$Te samples with Mn
content $y$$\,$$\leq$$\,$0.039 the observed behavior is similar to
the one observed in many canonical spin-glass systems with
predominance of the magnetic interactions with the exchange constant
$J$$\,$$>$$\,$0 [\onlinecite{Sereni93a}]. It is well known that in
case of a spin-glass systems the $M^{2}(B/M)$ curves on the Arrot
plot lies on the bottom side of the plot what reflects the absence
of spontaneous magnetization in the system. With the increase of the
amount of Mn above $y$$\,$$=$$\,$0.039 the $M^{2}(B/M)$ curves on
the Arrot plot shifts towards the top-left side of the plot. This is
a signature that the ferromagnetic interaction dominates in the
studied samples [\onlinecite{Sereni93a}]. Nevertheless, the system
remains in the frustrated spin-glass state (see
Fig.$\;$\ref{fig06}).

\section{Summary and conclusions}

We presented detailed studies of the magnetic properties of the
Ge$_{1\textrm{-}x\textrm{-}y}$Sn$_{x}$Mn$_{y}$Te crystals with
different amount of both magnetic and nonmagnetic constituents i.e.
Mn and Sn ions. \\ \indent The DC magnetometry measurements showed
that the studied spin-glass system shows some features
characteristic of a ferromagnetic material. Hysteresis loops with
large coercive fields (maximum of 500$\;$Oe) and spontaneous
magnetization were observed indicating that the studied system
possessed features characteristic of a ferromagnetic material. The
high field magnetization curves showed nonsaturating behavior at
magnetic fields as high as 90$\;$kOe which was interpreted as a
direct proof of a strong magnetic frustration in this system. The
observed bifurcations of the FC/ZFC magnetization $M(T)$
dependencies as well as the non-saturating $M(B)$ curves are the
features characteristic for spin-glasses. The experimental $M(T)$
curves were reproduced theoretically using the non-interacting spin
wave theory. The $M(B)$ curves were fitted theoretically using the
molecular field approximation. Theoretical calculations reproduced
well the experimental results especially in the range of high
magnetic fields. The linear $M(B)$ dependence was reproduced by
including magnetic disorder in the theoretical model and separate
treatment of the antiferromagnetically-coupled pairs of magnetic
ions. The proposed model gives satisfactory agreement with the
experimental curves only in the case of the presence of the
antiferromagnetic nearest-neighbor interactions in the system. This
is an evidence that strong magnetic disorder is present in the
studied system. \\ \indent The calculations based on a modified
Sherrington--Southern model showed that the spin glass state is the
preferred magnetic state for the experimentally determined hole
concentrations as well as Mn molar fractions. The calculations
allowed us to reproduce with good accuracy the spin glass transition
temperatures for all the studied crystals and to estimate the
Mn-hole exchange integral value to be around
$J_{pd}$$\,$$=$$\,$0.45$\pm$0.05$\;$eV for the studied
Ge$_{1\textrm{-}x\textrm{-}y}$Sn$_{x}$Mn$_{y}$Te mixed crystals.

\section{Acknowledgments}

\noindent The authors would like to thank Prof.~T.~Balcerzak for
helpful discussions. \\ \noindent This work was supported by the
Ministry of Science and Higher Education of Poland from a funds for
science in years 2009-2010 as a scientific project no. N N202
236537.

\appendix*
\section{Theoretical model including antiferromagnetic interactions}

The hamiltonian of the system of interacting magnetic ions
distributed over the lattice sites is:
\begin{equation}\label{EqHamiltonian}
\mathcal{H}=-\frac{1}{2}\sum_{i,j}^{}{J\left(R_{ij}\right)\xi_i\,\xi_j\,\mathbf{S}_i\,\mathbf{S}_j}-g_{S} \mu_{B} B\sum_{i}\xi_i\,\mathbf{S}_i^z,
\end{equation}
where the operators $\mathbf{S}_{i}$ describe the spins $S=5/2$
distributed over $N$ fcc lattice sites (the rhombohedral structure
of Ge$_{1\textrm{-}x\textrm{-}y}$Sn$_{x}$Mn$_{y}$ is very close to
the fcc structure). The exchange integral is given by
Eq.~(\ref{Eq03}). The disorder in the site-diluted magnetic system
is introduced by occupation number operators $\xi_i$ taking the
values of $0,1$ for each lattice site $i$. The set of $\xi_i$ values
describes completely the given realization of the disorder in the
system. \\ \indent Because of presence of the disorder, the lattice
sites occupied by magnetic ions are inequivalent in the
thermodynamic meaning. The most pronounced source of inhomogenity is
due to additional superexchange coupling occurring between the
magnetic ions in nearest-neighbor position. In order to account for
this, the system of magnetic impurities located on lattice sites
(with atomic concentration equal to $y$) was subdivided into two
(interacting) subsystems. For low values of $y$ the significant
number of impurity ions can be expected to lack the other magnetic
impurities in the nearest-neighbor position. Therefore, such
magnetic ions do not couple antiferromagnetically to other magnetic
ions via the superexchange mechanism. Let us denote the total number
of such ions (called further 'isolated magnetic ions') by $Npy$. For
purely random, uncorrelated occupation of fcc lattice sites by
impurities, the probability that a selected magnetic ion is isolated
equals to $p=\left(1-y\right)^{12}$ (what yields the value of $0.66$
for $y=0.034$). However, if the distribution of magnetic impurities
is not completely random, $p$ deviates from the above mentioned
value ($p<\left(1-y\right)^{12}$ for ion clustering
tendency, $p>\left(1-y\right)^{12}$ for the opposite tendency). \\
\indent We assume approximately that the remaining magnetic ions
form isolated nearest-neighbor pairs (i.e. pairs of ions that do not
possess further magnetic nearest neighbors). Within each pair, both
ions are coupled via the superexchange mechanism, characterized by
the exchange integral $J_{AF}<0$. We neglect the possibility of
formation of magnetic clusters containing more than two
superexchange-coupled magnetic ions, which is justified for small
$y$. According to this assumption, the number of isolated pairs is
$\frac{1}{2}y\left(1-p\right)$. Let us emphasize that due to the
neglect of larger magnetic clusters, the value of parameter $p$
determined from the best-fit to the experimental data may differ
from the
actual fraction of isolated magnetic ions in the real system. \\
\indent To construct the thermodynamic description of the model, we
apply a molecular-field approximation, assuming that the total state
of the system takes the form of a tensor product of the appropriate
single-site density matrices. The quantum state of each site
occupied by an isolated magnetic ion is described by a density
matrix of the form
\begin{equation}\label{EqRho0}
\boldsymbol\rho_{i,0}=\exp\left[\left(\lambda_{0}+g_{S} \mu_{B} B\right)\mathbf{S}_i^z/k_{\rm B}T\right]/Z_{0},
\end{equation}
with $Z_{0}=\sum_{s=-S}^{S}{\exp\left[\left(\lambda_{0}+g_{S}
\mu_{B} B\right)s/k_{\rm B}T\right]}$. The thermodynamic average of
the spin value of such an ion equals to
\begin{equation}\label{EqM0}
\left\langle \mathbf{S}^{z}_{i}\right\rangle=m_{0}=S\mathcal{B}_{S}\left(\frac{\lambda_{0}+g_{S} \mu_{B} B}{k_{\rm B}T}\right),
\end{equation}
$S\mathcal{B}_{S}\left(x\right)$ being the Brillouin function for
spin $S$. \\ \indent Let us further denote one of the (inequivalent)
ions belonging to an isolated nearest-neighbor pair by $+$, while
the other one by $-$. The corresponding density matrices are assumed
in the form:
\begin{equation}\label{EqRhoplusminus}
\boldsymbol\rho_{i,\pm}=\exp\left[\left(\lambda_{\pm}+g_{S} \mu_{B} B\right)\mathbf{S}_i^z/k_{\rm B}T\right]/Z_{\pm},
\end{equation}
with $Z_{\pm}=\sum_{s=-S}^{S}{\exp\left[\left(\lambda_{\pm}+g_{S} \mu_{B} B\right)s/k_{\rm B}T\right]}$. The thermodynamic average of spin value of each of the ions is
\begin{equation}\label{EqMplusminus}
\left\langle \mathbf{S}^{z}_{i}\right\rangle=m_{\pm}=S\mathcal{B}_{S}\left(\frac{\lambda_{\pm}+g_{S} \mu_{B} B}{k_{\rm B}T}\right).
\end{equation}
The parameters $\lambda_{0}$, $\lambda_{+}$ and $\lambda_{-}$ are
the variational parameters of the molecular field. \\ \indent For
such a model, the total entropy is a sum of contributions
originating from each single isolated magnetic ion ($-k_{\rm
B}\mathrm{Tr}\,\left[\boldsymbol \rho_{i,0}\ln
\boldsymbol\rho_{i,0}\right]$) and from each isolated magnetic pair
($-k_{\rm B}\mathrm{Tr}\,\left[\boldsymbol\rho_{i,+}\ln
\boldsymbol\rho_{i,+}\right]-k_{\rm
B}\mathrm{Tr}\,\left[\boldsymbol\rho_{i,-}\ln
\boldsymbol\rho_{i,-}\right]$). The total entropy of the system is:
\begin{widetext}
\begin{eqnarray}\label{EqEntropy}
\mathcal{S}=k_{\rm B}\sum_{i}^{}{\xi_{i}\,\left[ p\ln Z_{i,0}+\frac{1-p}{2}\left(\ln Z_{i,+}+\ln Z_{i,-}\right)\right.}+\nonumber\\-\left.p\frac{\lambda_{0}+g_{S} \mu_{B} B}{k_{\rm B}T}m_{0}- \frac{1-p}{2}\left(\frac{\lambda_{+}+g_{S} \mu_{B} B}{k_{\rm B}T}m_{+}+\frac{\lambda_{-}+g_{S} \mu_{B} B}{k_{\rm B}T}m_{-}\right)\right].
\end{eqnarray}
\end{widetext}

The enthalpy of the model is a thermodynamic average of the
Hamiltonian, containing the terms describing the interaction between
the isolated ions, between the isolated magnetic pairs as well as
the interaction between isolated magnetic ions and isolated pairs.
The total enthalpy amounts to:

\begin{widetext}
\begin{eqnarray}\label{EqEnthalpy}
\langle \mathcal{H}\rangle&=&-\sum_{i}^{}\xi_{i}\,\Bigg{\{}\frac{1}{2}p^2m_{0}^{2}\sum_{j}^{}{\xi_{j}J(R_{ij})}
+pm_{0}g_{S} \mu_{B} B+\frac{1}{2}(1-p)J_{AF}m_{+}m_{-}+\nonumber\\ &+&\frac{1-p}{2}g_{S} \mu_{B} B(m_{+}+m_{-})+p\frac{1-p}{2}m_{0}(m_{+}+m_{-})\sum_{j}^{}\xi_{j}J(R_{ij})+
\nonumber\\ &+&\frac{1}{4}(1-p)^2\sum_{j}^{}{\xi_{j}J(R_{ij})(m_{+}+m_{-})^2}\Bigg{\}}.
\end{eqnarray}
\end{widetext}

Let us note that in the formula above, as well as in the rest of the
Appendix, the nearest-neighbors of the lattice site $i$ will be
excluded from summation over $j$.

The total Gibbs free energy of the system is obtained from the
expression $G=\left\langle \mathcal{H} \right\rangle-\mathcal{S}T$.
The variational minimization of the Gibbs energy with respect to
molecular field parameters yields the set of coupled self-consistent
equation of the form:
\begin{eqnarray}\label{EqEquationsforLambda}
&&\sum_{i}^{}{\xi_{i}\left\{\lambda_{0}-\left[pm_{0}+\frac{1-p}{2}\left(m_{+}+m_{-}\right)\right]\sum_{j}^{}{\xi_{j}J\left(R_{ij}\right)}\right\}}=0\nonumber\\
&&\sum_{i}^{}{\xi_{i}\left\{\lambda_{+}-J_{AF}m_{-} -pm_{0} \sum_{j}^{}{\xi_{j}J\left(R_{ij}\right)}\right\}}=0\nonumber\\
&&\sum_{i}^{}{\xi_{i}\left\{\lambda_{-}-J_{AF}m_{+} -pm_{0} \sum_{j}^{}{\xi_{j}J\left(R_{ij}\right)}\right\}}=0.
\end{eqnarray}

Then we are looking for the solution when all the terms under the
sum over $i$ vanish simultaneously, i.e. we obtain the following set
of equations :
\begin{eqnarray}\label{EqEquationsforLambda2}
&&\lambda_{0}=\left[pm_{0}+\frac{1-p}{2}\left(m_{+}+m_{-}\right)\right]\sum_{j}^{}{\xi_{j}J\left(R_{j}\right)}\nonumber\\
&&\lambda_{+}=J_{AF}m_{-} +pm_{0} \sum_{j}^{}{\xi_{j}J\left(R_{j}\right)}\nonumber\\
&&\lambda_{-}=J_{AF}m_{+} +pm_{0} \sum_{j}^{}{\xi_{j}J\left(R_{j}\right)},
\end{eqnarray}

which need to be solved together with the conditions (\ref{EqM0})
and (\ref{EqMplusminus}). The solution allows us to calculate the
total magnetization in the system for a given realization of
disorder as:

\begin{equation}\label{EqTotalMagnetization}
M=n_{m} g_{S} \mu_{B} \left[pm_{0}+\frac{1-p}{2}\left(m_{+}+m_{-}\right)\right].
\end{equation}

Finally, to obtain the disorder-averaged site magnetization, we
average the values obtained from the
Eq.~(\ref{EqTotalMagnetization}) for a sufficiently large number of
realizations of disorder. Each realization of disorder is simulated
numerically, by allowing the parameters $\xi_j$ to take the values
of $0,1$ according to the probability distribution
$p\left(\xi_j\right)=y\,\delta\left(\xi_j-1\right)+\left(1-y\right)\delta\left(\xi_j\right)$,
with the total concentration of magnetic impurities equal to $y$.

\bibliography{bib01}

\begin{thebibliography}{28}%
\makeatletter
\providecommand \@ifxundefined [1]{%
 \@ifx{#1\undefined}
}%
\providecommand \@ifnum [1]{%
 \ifnum #1\expandafter \@firstoftwo
 \else \expandafter \@secondoftwo
 \fi
}%
\providecommand \@ifx [1]{%
 \ifx #1\expandafter \@firstoftwo
 \else \expandafter \@secondoftwo
 \fi
}%
\providecommand \natexlab [1]{#1}%
\providecommand \enquote  [1]{``#1''}%
\providecommand \bibnamefont  [1]{#1}%
\providecommand \bibfnamefont [1]{#1}%
\providecommand \citenamefont [1]{#1}%
\providecommand \href@noop [0]{\@secondoftwo}%
\providecommand \href [0]{\begingroup \@sanitize@url \@href}%
\providecommand \@href[1]{\@@startlink{#1}\@@href}%
\providecommand \@@href[1]{\endgroup#1\@@endlink}%
\providecommand \@sanitize@url [0]{\catcode `\\12\catcode `\$12\catcode
  `\&12\catcode `\#12\catcode `\^12\catcode `\_12\catcode `\%12\relax}%
\providecommand \@@startlink[1]{}%
\providecommand \@@endlink[0]{}%
\providecommand \url  [0]{\begingroup\@sanitize@url \@url }%
\providecommand \@url [1]{\endgroup\@href {#1}{\urlprefix }}%
\providecommand \urlprefix  [0]{URL }%
\providecommand \Eprint [0]{\href }%
\providecommand \doibase [0]{http://dx.doi.org/}%
\providecommand \selectlanguage [0]{\@gobble}%
\providecommand \bibinfo  [0]{\@secondoftwo}%
\providecommand \bibfield  [0]{\@secondoftwo}%
\providecommand \translation [1]{[#1]}%
\providecommand \BibitemOpen [0]{}%
\providecommand \bibitemStop [0]{}%
\providecommand \bibitemNoStop [0]{.\EOS\space}%
\providecommand \EOS [0]{\spacefactor3000\relax}%
\providecommand \BibitemShut  [1]{\csname bibitem#1\endcsname}%
\let\auto@bib@innerbib\@empty
\bibitem [{\citenamefont {Fukuma}\ \emph {et~al.}(2002)\citenamefont {Fukuma},
  \citenamefont {Asada}, \citenamefont {Arifuku},\ and\ \citenamefont
  {Koyanagi}}]{Fukuma02a}%
  \BibitemOpen
  \bibfield  {author} {\bibinfo {author} {\bibfnamefont {Y.}~\bibnamefont
  {Fukuma}}, \bibinfo {author} {\bibfnamefont {H.}~\bibnamefont {Asada}},
  \bibinfo {author} {\bibfnamefont {M.}~\bibnamefont {Arifuku}}, \ and\
  \bibinfo {author} {\bibfnamefont {T.}~\bibnamefont {Koyanagi}},\ }\href@noop
  {} {\bibfield  {journal} {\bibinfo  {journal} {Appl. Phys. Lett.}\ }\textbf
  {\bibinfo {volume} {80}},\ \bibinfo {pages} {1013} (\bibinfo {year}
  {2002})}\BibitemShut {NoStop}%
\bibitem [{\citenamefont {Rodot}\ \emph {et~al.}(1966)\citenamefont {Rodot},
  \citenamefont {Lewis}, \citenamefont {Rodot}, \citenamefont {Villers},
  \citenamefont {Cohen},\ and\ \citenamefont {Mollard}}]{Rodot66a}%
  \BibitemOpen
  \bibfield  {author} {\bibinfo {author} {\bibfnamefont {M.}~\bibnamefont
  {Rodot}}, \bibinfo {author} {\bibfnamefont {J.}~\bibnamefont {Lewis}},
  \bibinfo {author} {\bibfnamefont {H.}~\bibnamefont {Rodot}}, \bibinfo
  {author} {\bibfnamefont {G.}~\bibnamefont {Villers}}, \bibinfo {author}
  {\bibfnamefont {J.}~\bibnamefont {Cohen}}, \ and\ \bibinfo {author}
  {\bibfnamefont {P.}~\bibnamefont {Mollard}},\ }\href@noop {} {\bibfield
  {journal} {\bibinfo  {journal} {J. Phys. Soc. Japan Suppl.}\ }\textbf
  {\bibinfo {volume} {21}},\ \bibinfo {pages} {627} (\bibinfo {year}
  {1966})}\BibitemShut {NoStop}%
\bibitem [{\citenamefont {Cochrane}\ \emph {et~al.}(1974)\citenamefont
  {Cochrane}, \citenamefont {Plischke},\ and\ \citenamefont
  {T\"{o}in-Olsen}}]{Cochrane74a}%
  \BibitemOpen
  \bibfield  {author} {\bibinfo {author} {\bibfnamefont {R.~W.}\ \bibnamefont
  {Cochrane}}, \bibinfo {author} {\bibfnamefont {M.}~\bibnamefont {Plischke}},
  \ and\ \bibinfo {author} {\bibfnamefont {J.~O.}\ \bibnamefont
  {T\"{o}in-Olsen}},\ }\href@noop {} {\bibfield  {journal} {\bibinfo  {journal}
  {Phys. Rev. B}\ }\textbf {\bibinfo {volume} {9}},\ \bibinfo {pages} {3013}
  (\bibinfo {year} {1974})}\BibitemShut {NoStop}%
\bibitem [{\citenamefont {Fukuma}\ \emph {et~al.}(2006)\citenamefont {Fukuma},
  \citenamefont {Sato}, \citenamefont {Fujimoto}, \citenamefont {Tsuji},
  \citenamefont {Kimura}, \citenamefont {Taniguchi}, \citenamefont {Senba},
  \citenamefont {Tanaka}, \citenamefont {Asada},\ and\ \citenamefont
  {Koyanagi}}]{Fukuma06a}%
  \BibitemOpen
  \bibfield  {author} {\bibinfo {author} {\bibfnamefont {Y.}~\bibnamefont
  {Fukuma}}, \bibinfo {author} {\bibfnamefont {H.}~\bibnamefont {Sato}},
  \bibinfo {author} {\bibfnamefont {K.}~\bibnamefont {Fujimoto}}, \bibinfo
  {author} {\bibfnamefont {K.}~\bibnamefont {Tsuji}}, \bibinfo {author}
  {\bibfnamefont {A.}~\bibnamefont {Kimura}}, \bibinfo {author} {\bibfnamefont
  {M.}~\bibnamefont {Taniguchi}}, \bibinfo {author} {\bibfnamefont
  {S.}~\bibnamefont {Senba}}, \bibinfo {author} {\bibfnamefont
  {A.}~\bibnamefont {Tanaka}}, \bibinfo {author} {\bibfnamefont
  {H.}~\bibnamefont {Asada}}, \ and\ \bibinfo {author} {\bibfnamefont
  {T.}~\bibnamefont {Koyanagi}},\ }\href@noop {} {\bibfield  {journal}
  {\bibinfo  {journal} {J. Appl. Phys.}\ }\textbf {\bibinfo {volume} {99}},\
  \bibinfo {pages} {08D510} (\bibinfo {year} {2006})}\BibitemShut {NoStop}%
\bibitem [{\citenamefont {Chen}\ \emph {et~al.}(2006)\citenamefont {Chen},
  \citenamefont {Teo}, \citenamefont {Jalil},\ and\ \citenamefont
  {Liew}}]{Chen06a}%
  \BibitemOpen
  \bibfield  {author} {\bibinfo {author} {\bibfnamefont {W.~Q.}\ \bibnamefont
  {Chen}}, \bibinfo {author} {\bibfnamefont {K.~L.}\ \bibnamefont {Teo}},
  \bibinfo {author} {\bibfnamefont {M.~B.~A.}\ \bibnamefont {Jalil}}, \ and\
  \bibinfo {author} {\bibfnamefont {T.}~\bibnamefont {Liew}},\ }\href@noop {}
  {\bibfield  {journal} {\bibinfo  {journal} {J. Appl. Phys.}\ }\textbf
  {\bibinfo {volume} {99}},\ \bibinfo {pages} {08D515} (\bibinfo {year}
  {2006})}\BibitemShut {NoStop}%
\bibitem [{\citenamefont {Lechner}\ \emph {et~al.}(2007)\citenamefont
  {Lechner}, \citenamefont {Kirchschlager}, \citenamefont {Springholz},
  \citenamefont {Schwarzl},\ and\ \citenamefont {Bauer}}]{Lechner07a}%
  \BibitemOpen
  \bibfield  {author} {\bibinfo {author} {\bibfnamefont {R.~T.}\ \bibnamefont
  {Lechner}}, \bibinfo {author} {\bibfnamefont {R.}~\bibnamefont
  {Kirchschlager}}, \bibinfo {author} {\bibfnamefont {G.}~\bibnamefont
  {Springholz}}, \bibinfo {author} {\bibfnamefont {T.}~\bibnamefont
  {Schwarzl}}, \ and\ \bibinfo {author} {\bibfnamefont {G.}~\bibnamefont
  {Bauer}},\ }in\ \href@noop {} {\emph {\bibinfo {booktitle} {Narrow Gap
  Semiconductors 13 url:
  http://www.ati.surrey.ac.uk/NGS13/presentations/MO3\_4.kpdf}}}\ (\bibinfo
  {year} {2007})\BibitemShut {NoStop}%
\bibitem [{\citenamefont {Knoff}\ \emph {et~al.}(2008)\citenamefont {Knoff},
  \citenamefont {Domukhovski}, \citenamefont {Dybko}, \citenamefont {Dziawa},
  \citenamefont {Gorska}, \citenamefont {Jakiela}, \citenamefont {Lusakowska},
  \citenamefont {Reszka}, \citenamefont {Taliashvili}, \citenamefont {Story},
  \citenamefont {Anderson},\ and\ \citenamefont {Rotundu}}]{Knoff08a}%
  \BibitemOpen
  \bibfield  {author} {\bibinfo {author} {\bibfnamefont {W.}~\bibnamefont
  {Knoff}}, \bibinfo {author} {\bibfnamefont {V.}~\bibnamefont {Domukhovski}},
  \bibinfo {author} {\bibfnamefont {K.}~\bibnamefont {Dybko}}, \bibinfo
  {author} {\bibfnamefont {P.}~\bibnamefont {Dziawa}}, \bibinfo {author}
  {\bibfnamefont {M.}~\bibnamefont {Gorska}}, \bibinfo {author} {\bibfnamefont
  {R.}~\bibnamefont {Jakiela}}, \bibinfo {author} {\bibfnamefont
  {E.}~\bibnamefont {Lusakowska}}, \bibinfo {author} {\bibfnamefont
  {A.}~\bibnamefont {Reszka}}, \bibinfo {author} {\bibfnamefont
  {B.}~\bibnamefont {Taliashvili}}, \bibinfo {author} {\bibfnamefont
  {T.}~\bibnamefont {Story}}, \bibinfo {author} {\bibfnamefont {J.~R.}\
  \bibnamefont {Anderson}}, \ and\ \bibinfo {author} {\bibfnamefont {C.~R.}\
  \bibnamefont {Rotundu}},\ }\href@noop {} {\bibfield  {journal} {\bibinfo
  {journal} {Acta Phys. Pol. A}\ }\textbf {\bibinfo {volume} {114}},\ \bibinfo
  {pages} {1159} (\bibinfo {year} {2008})}\BibitemShut {NoStop}%
\bibitem [{\citenamefont {Lim}\ \emph {et~al.}(2009)\citenamefont {Lim},
  \citenamefont {Bi}, \citenamefont {Teo}, \citenamefont {Feng}, \citenamefont
  {Liew},\ and\ \citenamefont {Chong}}]{Lim09a}%
  \BibitemOpen
  \bibfield  {author} {\bibinfo {author} {\bibfnamefont {S.~T.}\ \bibnamefont
  {Lim}}, \bibinfo {author} {\bibfnamefont {J.~F.}\ \bibnamefont {Bi}},
  \bibinfo {author} {\bibfnamefont {K.~L.}\ \bibnamefont {Teo}}, \bibinfo
  {author} {\bibfnamefont {Y.~P.}\ \bibnamefont {Feng}}, \bibinfo {author}
  {\bibfnamefont {T.}~\bibnamefont {Liew}}, \ and\ \bibinfo {author}
  {\bibfnamefont {T.~C.}\ \bibnamefont {Chong}},\ }\href@noop {} {\bibfield
  {journal} {\bibinfo  {journal} {Appl. Phys. Lett.}\ }\textbf {\bibinfo
  {volume} {95}},\ \bibinfo {pages} {072510} (\bibinfo {year}
  {2009})}\BibitemShut {NoStop}%
\bibitem [{\citenamefont {Kowalski}\ \emph {et~al.}(2010)\citenamefont
  {Kowalski}, \citenamefont {Pietrzyk}, \citenamefont {Knoff}, \citenamefont
  {{\L}usakowski}, \citenamefont {Sadowski}, \citenamefont {Adell},\ and\
  \citenamefont {Story}}]{Kowalski10a}%
  \BibitemOpen
  \bibfield  {author} {\bibinfo {author} {\bibfnamefont {B.~J.}\ \bibnamefont
  {Kowalski}}, \bibinfo {author} {\bibfnamefont {M.~A.}\ \bibnamefont
  {Pietrzyk}}, \bibinfo {author} {\bibfnamefont {W.}~\bibnamefont {Knoff}},
  \bibinfo {author} {\bibfnamefont {A.}~\bibnamefont {{\L}usakowski}}, \bibinfo
  {author} {\bibfnamefont {J.}~\bibnamefont {Sadowski}}, \bibinfo {author}
  {\bibfnamefont {J.}~\bibnamefont {Adell}}, \ and\ \bibinfo {author}
  {\bibfnamefont {T.}~\bibnamefont {Story}},\ }\href@noop {} {\bibfield
  {journal} {\bibinfo  {journal} {Physics Procedia}\ }\textbf {\bibinfo
  {volume} {3}},\ \bibinfo {pages} {1357} (\bibinfo {year} {2010})}\BibitemShut
  {NoStop}%
\bibitem [{\citenamefont {Kilanski}\ \emph {et~al.}(2008)\citenamefont
  {Kilanski}, \citenamefont {Arciszewska}, \citenamefont {Domukhovsky},
  \citenamefont {Dobrowolski}, \citenamefont {Slynko},\ and\ \citenamefont
  {Slynko}}]{Kilanski08a}%
  \BibitemOpen
  \bibfield  {author} {\bibinfo {author} {\bibfnamefont {L.}~\bibnamefont
  {Kilanski}}, \bibinfo {author} {\bibfnamefont {M.}~\bibnamefont
  {Arciszewska}}, \bibinfo {author} {\bibfnamefont {V.}~\bibnamefont
  {Domukhovsky}}, \bibinfo {author} {\bibfnamefont {W.}~\bibnamefont
  {Dobrowolski}}, \bibinfo {author} {\bibfnamefont {V.~E.}\ \bibnamefont
  {Slynko}}, \ and\ \bibinfo {author} {\bibfnamefont {E.~I.}\ \bibnamefont
  {Slynko}},\ }\href@noop {} {\bibfield  {journal} {\bibinfo  {journal} {Acta
  Phys. Pol. A}\ }\textbf {\bibinfo {volume} {114}},\ \bibinfo {pages} {1145}
  (\bibinfo {year} {2008})}\BibitemShut {NoStop}%
\bibitem [{\citenamefont {Kilanski}\ \emph {et~al.}(2009)\citenamefont
  {Kilanski}, \citenamefont {Arciszewska}, \citenamefont {Domukhovsky},
  \citenamefont {Dobrowolski}, \citenamefont {Slynko},\ and\ \citenamefont
  {Slynko}}]{Kilanski09a}%
  \BibitemOpen
  \bibfield  {author} {\bibinfo {author} {\bibfnamefont {L.}~\bibnamefont
  {Kilanski}}, \bibinfo {author} {\bibfnamefont {M.}~\bibnamefont
  {Arciszewska}}, \bibinfo {author} {\bibfnamefont {V.}~\bibnamefont
  {Domukhovsky}}, \bibinfo {author} {\bibfnamefont {W.}~\bibnamefont
  {Dobrowolski}}, \bibinfo {author} {\bibfnamefont {V.~E.}\ \bibnamefont
  {Slynko}}, \ and\ \bibinfo {author} {\bibfnamefont {E.~I.}\ \bibnamefont
  {Slynko}},\ }\href@noop {} {\bibfield  {journal} {\bibinfo  {journal} {J.
  Appl. Phys.}\ }\textbf {\bibinfo {volume} {105}},\ \bibinfo {pages} {103901}
  (\bibinfo {year} {2009})}\BibitemShut {NoStop}%
\bibitem [{\citenamefont {Aust}\ and\ \citenamefont
  {Chalmers}(1958)}]{Aust58a}%
  \BibitemOpen
  \bibfield  {author} {\bibinfo {author} {\bibfnamefont {K.~T.}\ \bibnamefont
  {Aust}}\ and\ \bibinfo {author} {\bibfnamefont {B.}~\bibnamefont
  {Chalmers}},\ }\href@noop {} {\bibfield  {journal} {\bibinfo  {journal} {Can.
  J. Phys.}\ }\textbf {\bibinfo {volume} {36}},\ \bibinfo {pages} {977}
  (\bibinfo {year} {1958})}\BibitemShut {NoStop}%
\bibitem [{\citenamefont {Galazka}\ \emph {et~al.}(1999)\citenamefont
  {Galazka}, \citenamefont {Kossut},\ and\ \citenamefont {Story}}]{Galazka99a}%
  \BibitemOpen
  \bibfield  {author} {\bibinfo {author} {\bibfnamefont {R.~R.}\ \bibnamefont
  {Galazka}}, \bibinfo {author} {\bibfnamefont {J.}~\bibnamefont {Kossut}}, \
  and\ \bibinfo {author} {\bibfnamefont {T.}~\bibnamefont {Story}},\ }\enquote
  {\bibinfo {title} {Landolt-b\"{o}rnstein, new series, group iii/41},}\ \
  (\bibinfo  {publisher} {Berlin, Heidelberg, Springer-Verlag},\ \bibinfo
  {year} {1999})\ Chap.\ \bibinfo {chapter} {Semiconductors}\BibitemShut
  {NoStop}%
\bibitem [{\citenamefont {Mydosh}(1994)}]{Mydosh94a}%
  \BibitemOpen
  \bibfield  {author} {\bibinfo {author} {\bibfnamefont {J.~A.}\ \bibnamefont
  {Mydosh}},\ }\href@noop {} {\emph {\bibinfo {title} {Spin Glasses: An
  Experimental Introduction}}}\ (\bibinfo  {publisher} {Taylor and Francis,
  London},\ \bibinfo {year} {1994})\BibitemShut {NoStop}%
\bibitem [{\citenamefont {Ruderman}\ and\ \citenamefont
  {Kittel}(1954)}]{Ruderman54a}%
  \BibitemOpen
  \bibfield  {author} {\bibinfo {author} {\bibfnamefont {M.~A.}\ \bibnamefont
  {Ruderman}}\ and\ \bibinfo {author} {\bibfnamefont {C.}~\bibnamefont
  {Kittel}},\ }\href@noop {} {\bibfield  {journal} {\bibinfo  {journal} {Phys.
  Rev.}\ }\textbf {\bibinfo {volume} {96}},\ \bibinfo {pages} {99} (\bibinfo
  {year} {1954})}\BibitemShut {NoStop}%
\bibitem [{\citenamefont {Kasuya}(1956)}]{Kasuya56a}%
  \BibitemOpen
  \bibfield  {author} {\bibinfo {author} {\bibfnamefont {T.}~\bibnamefont
  {Kasuya}},\ }\href@noop {} {\bibfield  {journal} {\bibinfo  {journal} {Progr.
  Theor. Phys.}\ }\textbf {\bibinfo {volume} {16}},\ \bibinfo {pages} {45}
  (\bibinfo {year} {1956})}\BibitemShut {NoStop}%
\bibitem [{\citenamefont {Yoshida}(1957)}]{Yoshida57a}%
  \BibitemOpen
  \bibfield  {author} {\bibinfo {author} {\bibfnamefont {K.}~\bibnamefont
  {Yoshida}},\ }\href@noop {} {\bibfield  {journal} {\bibinfo  {journal} {Phys.
  Rev.}\ }\textbf {\bibinfo {volume} {106}},\ \bibinfo {pages} {893} (\bibinfo
  {year} {1957})}\BibitemShut {NoStop}%
\bibitem [{\citenamefont {Cochrane}\ \emph {et~al.}(1973)\citenamefont
  {Cochrane}, \citenamefont {Hedgcock},\ and\ \citenamefont
  {Str\"{o}m-Olsen}}]{Cochrane73a}%
  \BibitemOpen
  \bibfield  {author} {\bibinfo {author} {\bibfnamefont {R.~W.}\ \bibnamefont
  {Cochrane}}, \bibinfo {author} {\bibfnamefont {F.~T.}\ \bibnamefont
  {Hedgcock}}, \ and\ \bibinfo {author} {\bibfnamefont {J.~O.}\ \bibnamefont
  {Str\"{o}m-Olsen}},\ }\href@noop {} {\bibfield  {journal} {\bibinfo
  {journal} {Phys. Rev. B}\ }\textbf {\bibinfo {volume} {8}},\ \bibinfo {pages}
  {4262} (\bibinfo {year} {1973})}\BibitemShut {NoStop}%
\bibitem [{\citenamefont {Fukuma}\ \emph {et~al.}(2008)\citenamefont {Fukuma},
  \citenamefont {Asada}, \citenamefont {Miyawaki}, \citenamefont {Koyanagi},
  \citenamefont {Senba}, \citenamefont {Goto},\ and\ \citenamefont
  {Sato}}]{Fukuma08b}%
  \BibitemOpen
  \bibfield  {author} {\bibinfo {author} {\bibfnamefont {Y.}~\bibnamefont
  {Fukuma}}, \bibinfo {author} {\bibfnamefont {H.}~\bibnamefont {Asada}},
  \bibinfo {author} {\bibfnamefont {S.}~\bibnamefont {Miyawaki}}, \bibinfo
  {author} {\bibfnamefont {T.}~\bibnamefont {Koyanagi}}, \bibinfo {author}
  {\bibfnamefont {S.}~\bibnamefont {Senba}}, \bibinfo {author} {\bibfnamefont
  {K.}~\bibnamefont {Goto}}, \ and\ \bibinfo {author} {\bibfnamefont
  {H.}~\bibnamefont {Sato}},\ }\href@noop {} {\bibfield  {journal} {\bibinfo
  {journal} {Appl. Phys. Lett.}\ }\textbf {\bibinfo {volume} {93}},\ \bibinfo
  {pages} {252502} (\bibinfo {year} {2008})}\BibitemShut {NoStop}%
\bibitem [{\citenamefont {Sherrington}\ and\ \citenamefont
  {Southern}(1975)}]{Sherrington75b}%
  \BibitemOpen
  \bibfield  {author} {\bibinfo {author} {\bibfnamefont {D.}~\bibnamefont
  {Sherrington}}\ and\ \bibinfo {author} {\bibfnamefont {B.~W.}\ \bibnamefont
  {Southern}},\ }\href@noop {} {\bibfield  {journal} {\bibinfo  {journal} {J.
  Phys. F: Met. Phys.}\ }\textbf {\bibinfo {volume} {5}},\ \bibinfo {pages}
  {L49} (\bibinfo {year} {1975})}\BibitemShut {NoStop}%
\bibitem [{\citenamefont {Eggenkamp}\ \emph {et~al.}(1995)\citenamefont
  {Eggenkamp}, \citenamefont {Swagten}, \citenamefont {Story}, \citenamefont
  {Litvinov}, \citenamefont {Sw\"{u}ste},\ and\ \citenamefont
  {de~Jonge}}]{Eggenkamp95a}%
  \BibitemOpen
  \bibfield  {author} {\bibinfo {author} {\bibfnamefont {P.~J.~T.}\
  \bibnamefont {Eggenkamp}}, \bibinfo {author} {\bibfnamefont {H.~J.~M.}\
  \bibnamefont {Swagten}}, \bibinfo {author} {\bibfnamefont {T.}~\bibnamefont
  {Story}}, \bibinfo {author} {\bibfnamefont {V.~I.}\ \bibnamefont {Litvinov}},
  \bibinfo {author} {\bibfnamefont {C.~H.~W.}\ \bibnamefont {Sw\"{u}ste}}, \
  and\ \bibinfo {author} {\bibfnamefont {W.~J.~M.}\ \bibnamefont {de~Jonge}},\
  }\href@noop {} {\bibfield  {journal} {\bibinfo  {journal} {Phys. Rev. B}\
  }\textbf {\bibinfo {volume} {51}},\ \bibinfo {pages} {15250} (\bibinfo {year}
  {1995})}\BibitemShut {NoStop}%
\bibitem [{\citenamefont {Lewis}(1969)}]{Lewis69a}%
  \BibitemOpen
  \bibfield  {author} {\bibinfo {author} {\bibfnamefont {J.~E.}\ \bibnamefont
  {Lewis}},\ }\href@noop {} {\bibfield  {journal} {\bibinfo  {journal} {Phys.
  Stat. Sol. B}\ }\textbf {\bibinfo {volume} {35}},\ \bibinfo {pages} {737}
  (\bibinfo {year} {1969})}\BibitemShut {NoStop}%
\bibitem [{\citenamefont {Lewis}(1973)}]{Lewis73a}%
  \BibitemOpen
  \bibfield  {author} {\bibinfo {author} {\bibfnamefont {J.~E.}\ \bibnamefont
  {Lewis}},\ }\href@noop {} {\bibfield  {journal} {\bibinfo  {journal} {Phys.
  Stat. Sol. B}\ }\textbf {\bibinfo {volume} {59}},\ \bibinfo {pages} {367}
  (\bibinfo {year} {1973})}\BibitemShut {NoStop}%
\bibitem [{\citenamefont {Eggenkamp}\ \emph {et~al.}(1994)\citenamefont
  {Eggenkamp}, \citenamefont {Vennix}, \citenamefont {Story}, \citenamefont
  {Swagten}, \citenamefont {Sw\"{u}ste},\ and\ \citenamefont
  {de~Jonge}}]{Eggenkamp94a}%
  \BibitemOpen
  \bibfield  {author} {\bibinfo {author} {\bibfnamefont {P.~J.~T.}\
  \bibnamefont {Eggenkamp}}, \bibinfo {author} {\bibfnamefont {C.~W.~H.~M.}\
  \bibnamefont {Vennix}}, \bibinfo {author} {\bibfnamefont {T.}~\bibnamefont
  {Story}}, \bibinfo {author} {\bibfnamefont {H.~J.~M.}\ \bibnamefont
  {Swagten}}, \bibinfo {author} {\bibfnamefont {C.~H.~W.}\ \bibnamefont
  {Sw\"{u}ste}}, \ and\ \bibinfo {author} {\bibfnamefont {W.~J.~M.}\
  \bibnamefont {de~Jonge}},\ }\href@noop {} {\bibfield  {journal} {\bibinfo
  {journal} {J. Appl. Phys.}\ }\textbf {\bibinfo {volume} {75}},\ \bibinfo
  {pages} {5728} (\bibinfo {year} {1994})}\BibitemShut {NoStop}%
\bibitem [{\citenamefont {Fukuma}\ \emph {et~al.}(2003)\citenamefont {Fukuma},
  \citenamefont {Asada}, \citenamefont {Nishimura},\ and\ \citenamefont
  {Koyanagi}}]{Fukuma03a}%
  \BibitemOpen
  \bibfield  {author} {\bibinfo {author} {\bibfnamefont {Y.}~\bibnamefont
  {Fukuma}}, \bibinfo {author} {\bibfnamefont {H.}~\bibnamefont {Asada}},
  \bibinfo {author} {\bibfnamefont {N.}~\bibnamefont {Nishimura}}, \ and\
  \bibinfo {author} {\bibfnamefont {T.}~\bibnamefont {Koyanagi}},\ }\href@noop
  {} {\bibfield  {journal} {\bibinfo  {journal} {J. Appl. Phys.}\ }\textbf
  {\bibinfo {volume} {93}},\ \bibinfo {pages} {4034} (\bibinfo {year}
  {2003})}\BibitemShut {NoStop}%
\bibitem [{\citenamefont {Prejean}\ \emph {et~al.}(1980)\citenamefont
  {Prejean}, \citenamefont {Joliclerc},\ and\ \citenamefont
  {Monod}}]{Prejean80a}%
  \BibitemOpen
  \bibfield  {author} {\bibinfo {author} {\bibfnamefont {J.~J.}\ \bibnamefont
  {Prejean}}, \bibinfo {author} {\bibfnamefont {M.~J.}\ \bibnamefont
  {Joliclerc}}, \ and\ \bibinfo {author} {\bibfnamefont {P.}~\bibnamefont
  {Monod}},\ }\href@noop {} {\bibfield  {journal} {\bibinfo  {journal} {J.
  Phys. (Paris)}\ }\textbf {\bibinfo {volume} {41}},\ \bibinfo {pages} {427}
  (\bibinfo {year} {1980})}\BibitemShut {NoStop}%
\bibitem [{\citenamefont {Senoussi}(1984)}]{Senoussi84a}%
  \BibitemOpen
  \bibfield  {author} {\bibinfo {author} {\bibfnamefont {S.}~\bibnamefont
  {Senoussi}},\ }\href@noop {} {\bibfield  {journal} {\bibinfo  {journal} {J.
  Phys.}\ }\textbf {\bibinfo {volume} {45}},\ \bibinfo {pages} {315} (\bibinfo
  {year} {1984})}\BibitemShut {NoStop}%
\bibitem [{\citenamefont {Sereni}\ \emph {et~al.}(1993)\citenamefont {Sereni},
  \citenamefont {Beaurepaire},\ and\ \citenamefont {Kappler}}]{Sereni93a}%
  \BibitemOpen
  \bibfield  {author} {\bibinfo {author} {\bibfnamefont {J.~G.}\ \bibnamefont
  {Sereni}}, \bibinfo {author} {\bibfnamefont {E.}~\bibnamefont {Beaurepaire}},
  \ and\ \bibinfo {author} {\bibfnamefont {J.~P.}\ \bibnamefont {Kappler}},\
  }\href@noop {} {\bibfield  {journal} {\bibinfo  {journal} {Phys. Rev. B}\
  }\textbf {\bibinfo {volume} {48}},\ \bibinfo {pages} {3747} (\bibinfo {year}
  {1993})}\BibitemShut {NoStop}%
\end{thebibliography}%

\end{document}